\documentclass[journal,12pt,draftclsnofoot,onecolumn,twoside]{IEEEtran}
\usepackage[cmex10]{amsmath}
\usepackage{amssymb}

\usepackage{ifpdf}
 \ifpdf
     \usepackage[pdftex]{graphicx}
 \else
     \usepackage[dvips]{graphicx}
 \fi

\graphicspath{{figures/}}
\usepackage[footnotesize]{caption}
\usepackage{epstopdf}					

\usepackage{cite}
\usepackage{subfig}
\newtheorem{proposition}{Proposition}
\newcommand{\rttensor}[1]{\overline{\overline{#1}}}
\usepackage{algorithm}
\usepackage{algpseudocode}
\usepackage{flushend}

\ifCLASSINFOpdf
\else
\fi

\begin{document}

\title{Blind Identification of SM and Alamouti STBC-OFDM Signals}

\author{Yahia~A.~Eldemerdash,~\IEEEmembership{Student Member,~IEEE,}
        Octavia~A.~Dobre,~\IEEEmembership{Senior Member,~IEEE}, and  Bruce~J.~Liao,~\IEEEmembership{Senior Member,~IEEE}

\thanks{This work was supported in part by the Defence Research and Development
Canada (DRDC)}
     
\thanks{Yahia A. Eldemerdash, and  Octavia A. Dobre are with the Faculty of Engineering and Applied Science, Memorial
University of Newfoundland,  St. John's, Canada. 
Email: \{yahia.eldemerdash, odobre\}@mun.ca.}
\thanks{Bruce J. Liao is with Defence Research and Development Canada, Ottawa, Canada. E-mail:
bruce.liao@drdc-rddc.gc.ca.}
}

\maketitle

\vspace{-1.5cm}
\begin{abstract}
This paper proposes an efficient identification algorithm for spatial multiplexing (SM) and Alamouti (AL) coded orthogonal frequency division multiplexing (OFDM) signals. 
The cross-correlation between the received signals from different antennas is exploited to provide a discriminating feature to identify SM-OFDM and AL-OFDM signals. The proposed algorithm requires neither estimation of the channel coefficients and noise power, nor the modulation of the transmitted signal. Moreover, it does not need space-time block code (STBC) or OFDM block synchronization. The effectiveness of the proposed algorithm is demonstrated through extensive simulation experiments in the presence of diverse transmission impairments, such as time and frequency offsets, Doppler frequency, and spatially correlated fading.
\end{abstract}


%
\IEEEpeerreviewmaketitle

\section{Introduction}
\IEEEPARstart{B}{lind}
signal identification plays an important role in various military and commercial applications, including electronic warfare, radio surveillance, software defined radio, and spectrum awareness in cognitive radio \cite{dobre2007survey,cabric2008addressing,xu2010software}. 
For example, in software
defined radio the transmitter provides a  flexible architecture, in which the same hardware can be used for different transmission parameters, e.g., modulation format, coding rate, and antenna configuration. Accordingly,  algorithms are required at the receive-side to blindly estimate these signal parameters \cite{xu2010software}.

Numerous studies have addressed the problem of blind signal identification in single-input single-output scenarios. These include  identification  of the modulation format \cite{wu2008novel,su203MOD_class,su2008real,dobre2012cyclostationarity,grimaldi2007automatic},
single- versus multi-carrier transmissions \cite{zhang2013second}, the type of multi-carrier technique \cite{bouzegzi2010new,al2012second}, and channel encoders \cite{xia2013blind,xia2014novel,xia2014joint}, as well as blind parameter estimation \cite{wu2005pilot,zhang2013second}.
Recently, multiple-input multiple-output (MIMO) technology has been adopted 
by different wireless standards, such as IEEE 802.11n, IEEE 802.16e, and 3GPP LTE \cite{LTE_WiMAX}.
However, the study of MIMO signal identification is at an early stage. For example, estimation of the number of transmit antennas has been investigated in \cite{No_Trans_anrennas1,No_Trans_anrennas2}, modulation identification in \cite{hassan2012blind, MOD_MIMO_Choqueuse,muhlhaus203MOD_class}, and space-time block code (STBC) identification in \cite{choqueuse2008hierarchical,choqueuse2010blind,marey2012classification,luo2012blind,qian2013blind,yahia2013TCOM}. All these studies considered single-carrier transmission over frequency-flat fading.
However, in practice high data rate applications  necessitate transmissions over frequency-selective channels; hence, the assumption of frequency-flat fading is not practically accepted. 
Additionally, the orthogonal frequency division multiplexing (OFDM) technique
has been adopted as the main transmission scheme over frequency-selective fading channels \cite{LTE_WiMAX}. Therefore, investigating the problem of MIMO-OFDM signal identification becomes a practically required challenge.
Recently, this problem has been explored in \cite{agirman2011modulation,MareyICC2013,marey2014blind,karami2014identification}: modulation identification for spatial multiplexing (SM)-OFDM was studied in \cite{agirman2011modulation} and STBC-OFDM signal identification was considered in \cite{MareyICC2013,marey2014blind,karami2014identification}, with the latter being relevant for our work. The identification algorithm proposed in \cite{MareyICC2013,marey2014blind}
requires a large observation period to achieve a good identification performance and suffers from high sensitivity to frequency offset. In addition to these drawbacks, the  algorithm in \cite{karami2014identification} is applicable only for a reduced number of OFDM subcarriers.

In this paper, we propose an efficient algorithm to blindly identify Alamouti (AL)-OFDM and SM-OFDM signals\footnote{Note that we assume that the received signal is either AL-OFDM or SM-OFDM. The AL and SM STBCs are considered, as they are  commonly used in various wireless standards, such as IEEE 802.11n, IEEE 802.16e, and 3GPP LTE \cite{LTE_WiMAX}.}. A novel cross-correlation is defined for the received sequences with re-arranged blocks, which provides a powerful discriminating feature. Additionally, a novel criterion of decision is developed based on the statistical properties of the feature estimate. The proposed algorithm does not require information about the channel, modulation format, noise power, or timing synchronization. Moreover, it has the advantage of providing a good identification performance with a short observation  period and for various numbers of OFDM subcarriers, as well as of being relatively robust to the frequency offset.

The rest of this paper is organized as follows. Section \ref{sec:SIGNAL-MODEL} introduces the system model. Section \ref{sec:algorithm} describes the proposed identification algorithm. Simulation results are presented in Section \ref{sec:simulation}. Finally, concluding remarks are drawn in Section~\ref{sec:conclusion}.

\begin{figure}
\begin{centering}
\includegraphics[width=0.75\textwidth]{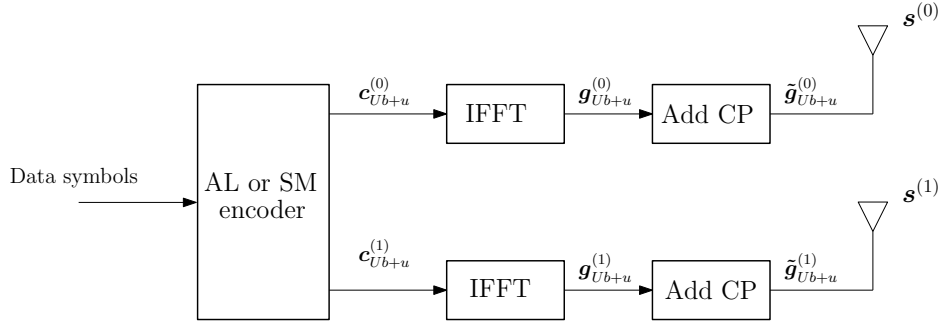}
\par\end{centering}
\caption{Block diagram of a MIMO-OFDM transmitter \cite{li2002mimo}.\label{fig:Block}}
\end{figure} 

\section{System Model\label{sec:SIGNAL-MODEL}}

We consider a MIMO-OFDM system with two transmit antennas, which employs either an AL or SM encoder, as shown in Fig. \ref{fig:Block}.
The data symbols, which are randomly and independently drawn from an $M$-point constellation, $M\geq 4$, are considered as blocks of length $N$. These are fed to the encoder, whose output is
$[\begin{array}{cc}
\boldsymbol{c}_{2b+0}^{(0)} & \boldsymbol{c}_{2b+1}^{(0)};
\end{array} 
\begin{array}{cc}
\boldsymbol{c}_{2b+0}^{(1)} & \boldsymbol{c}_{2b+1}^{(1)}
\end{array}]$
for AL-OFDM and 
$[\begin{array}{cc}
\boldsymbol{c}_{b+0}^{(0)}; & \boldsymbol{c}_{b+0}^{(1)}
\end{array}]$
for SM-OFDM. The notation $\boldsymbol{c}_{Ub+u}^{(f)}=[c_{Ub+u}^{(f)}(0),...,c_{Ub+u}^{(f)}(N-1)]$ is used to represent the $(Ub+u)$th data block of $N$ symbols from the $f$th antenna, $f=0,1$, with $b$ as the STBC block index, $U$ as the length of the STBC block ($U=2$ for AL and $U=1$ for SM), and $u$ as the slot index within an STBC block, $u=0,1,...,U-1$. For AL-OFDM, the data blocks have the property that  \cite{MareyICC2013}: $\boldsymbol{c}_{2b+1}^{(1)}=(\boldsymbol{c}_{2b+0}^{(0)})^*$ and $\boldsymbol{c}_{2b+1}^{(0)}=-(\boldsymbol{c}_{2b+0}^{(1)})^*$, where $*$ denotes complex conjugate.

Each block $\boldsymbol{c}_{Ub+u}^{(f)}$ is input to an $N$-point inverse fast Fourier transform ($N$-IFFT), leading to the time-domain block  $\boldsymbol{g}_{Ub+u}^{(f)}=[g_{Ub+u}^{(f)}(0),$  $g_{Ub+u}^{(f)}(1),...,g_{Ub+u}^{(f)}(N-1)]$. Then, a cyclic prefix of length $\nu$ is added, with the resulting  OFDM block written as $\tilde{\boldsymbol{g}}_{Ub+u}^{(f)}=[\tilde{g}_{Ub+u}^{(f)}(0),...,\tilde{g}_{Ub+u}^{(f)}(\nu),\tilde{g}_{Ub+u}^{(f)}(\nu$ $+1),...,\tilde{g}_{Ub+u}^{(f)}(N+\nu-1)]=[g_{Ub+u}^{(f)}(N-\nu),...,g_{Ub+u}^{(f)}(0),$ $g_{Ub+u}^{(f)}(1),...,g_{Ub+u}^{(f)}(N-1)]$. Accordingly, the time-domain samples of the OFDM block can be expressed as

\vspace{-0.25cm}

\begin{equation}
\begin{array}{l}
\tilde{g}_{Ub+u}^{(f)}(n)=\frac{1}{\sqrt{N}}\sum_{p=0}^{N-1}c_{Ub+u}^{(f)}(p)e^{\frac{j2\pi p(n-\nu)}{N}}, 
\quad n=0,1,..,N+\nu-1.\label{eq:IFFT_eq}
\end{array}
\end{equation}

With the transmit sequence from the $f$th antenna as $\boldsymbol{s}^{(f)}=\left[...\boldsymbol{\tilde{g}}_{-1}^{(f)},\boldsymbol{\tilde{g}}_{0}^{(f)},\boldsymbol{\tilde{g}}_{1}^{(f)},\boldsymbol{\tilde{g}}_{2}^{(f)},...\right]$, whose $k$th element is denoted by $s^{(f)}(k)$,
the $k$th received sample at the $i$th receive antenna, $i=0,1,...,N_r-1$, can be expressed as \cite{MareyICC2013}

\vspace{-0.25cm}
\begin{equation}
r^{(i)}(k)=\sum_{f=0}^{1}\sum_{l=0}^{L_{h}-1}h_{fi}(l)s^{(f)}(k-l)+w^{(i)}(k),\label{eq:RX_signal}\end{equation}
where $L_h$ is the number of propagation paths, $h_{fi}(l)$ is the channel coefficient corresponding to the $l$th path between the transmit antenna $f$ and the receive antenna $i$, and $w^{(i)}(k)$ represents the complex additive white Gaussian noise (AWGN) at the $i$th receive antenna, with zero mean and variance~$\sigma^2_w$.

\section{Proposed algorithm \label{sec:algorithm}}
In this section, we investigate the second-order cross-correlation as 
a discriminating feature  for AL-OFDM and SM-OFDM signal identification. Initially, we consider the $N_r=2$ case, for which we explore the cross-correlation between
$\{ r^{(0)}(k)\} $ and $\{ r^{(1)}(k)\} $ and  develop a new decision criterion based on the statistical properties of the feature estimate. Then, we extend the analysis to the case of $N_r>2$.

\vspace{-0.2cm}
\subsection{Cross-correlation properties ($N_r=2$)}

First, the cross-correlation properties for AL-OFDM and SM-OFDM signals are analyzed at the transmit-side, and then the analysis is extended at the receive-side.

\vspace{0.15cm}
\underline{\textit{Transmit-side}}

\begin{figure}
\begin{centering}
\includegraphics[width=0.75\textwidth]{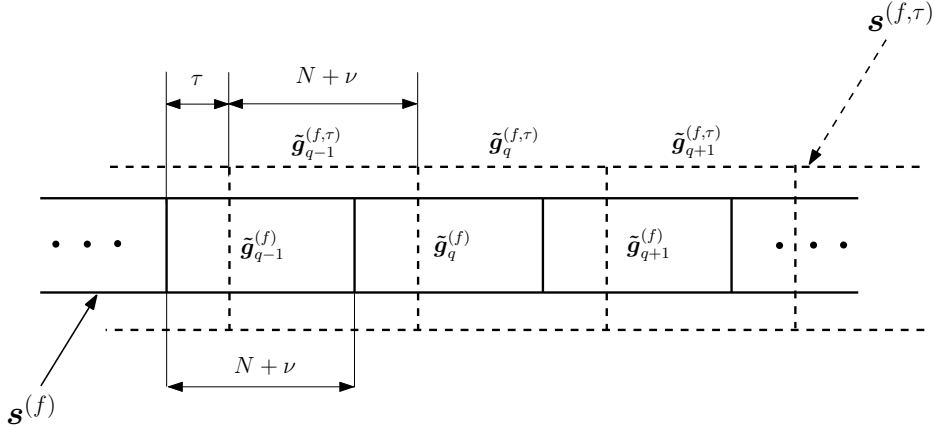}
\par\end{centering}
\caption{Illustration of the relation between the $\boldsymbol{s}^{(f)}$ and  $\boldsymbol{s}^{(f,\tau)}$ sequences. Solid lines are used to delimitate the OFDM blocks of $\boldsymbol{s}^{(f)}$, while dashed lines show the $(N+\nu)$-length blocks of   $\boldsymbol{s}^{(f,\tau)}$. \label{fig:g_delay}}
\end{figure}

Let us form the sequence $\boldsymbol{s}^{(f,\tau)}$, whose components are given by
 $s^{(f,\tau)}(k)=s^{(f)}(k+\tau)$,  $\tau=0,1,...,N+\nu-1$. This is further divided into consecutive ($N+\nu$)-length blocks, i.e., $\boldsymbol{s}^{(f,\tau)}= [...\boldsymbol{\tilde{g}}^{(f,\tau)}_{-1},\boldsymbol{\tilde{g}}^{(f,\tau)}_{0}, \boldsymbol{\tilde{g}}^{(f,\tau)}_{1},$ $...,\boldsymbol{\tilde{g}}^{(f,\tau)}_{q-1},\boldsymbol{\tilde{g}}^{(f,\tau)}_{q},\boldsymbol{\tilde{g}}^{(f,\tau)}_{q+1},...]$, as it is  graphically illustrated in Fig.~\ref{fig:g_delay}.

\vspace{0.2cm}
\begin{proposition}
For the AL-OFDM signal, the samples of the $(N+\nu)$-length blocks of the newly formed sequence 
$\boldsymbol{s}^{(f,\tau)}$ exhibits the following properties:

\begin{subequations} \label{eq:test_test}

\begin{equation}
\begin{array}{l}
 \hspace{-2.7cm} \bullet  \;\tau=0:   \tilde{g}_{2b+0}^{(0,0)}(n)=\tilde{g}_{2b+1}^{(1,0)^{*}}(mod(-(n- \nu),N)+ \nu),  \quad n=0,1,...,N+\nu-1,
\end{array}
\label{eq:G_0}
\end{equation}
\begin{equation}
\begin{array}{l}
 \hspace{-3.5cm}\bullet \;\tau=N/2: 
 
 \tilde{g}_{2b+0}^{(0,\frac{N}{2})}(n)=\tilde{g}_{2b+1}^{(1,\frac{N}{2})^{*}}(mod(-(n- \nu),N)+ \nu),  \quad n=0,1,...,\nu,
 
  \end{array} 
\label{eq:G_1}
\end{equation}
\begin{equation}
\begin{array}{l}
 \hspace{-0.371cm}\bullet \;\tau=N/2+\nu:  \tilde{g}_{2b-1}^{(0,\frac{N}{2}+\nu)}(n)=\tilde{g}_{2b+0}^{(1,\frac{N}{2}+\nu)^{*}}(mod(-(n- \nu),N)+ \nu),  n=\frac{N}{2},\frac{N}{2}+1,...,\frac{N}{2}+2\nu.
\end{array} 
\label{eq:G_2}
\end{equation}

\end{subequations}
Such properties do not hold for any other values of $\tau$ and $n$. Additionally, these are not valid for the SM-OFDM signal.

\end{proposition}
\vspace{0.35cm}
\begin{IEEEproof}
See Appendix.
\end{IEEEproof}

\begin{figure*}
  \captionsetup[subfigure]{labelformat=empty}
  \quad \quad \quad 
  \subfloat[]{\includegraphics[width=.25\textwidth]{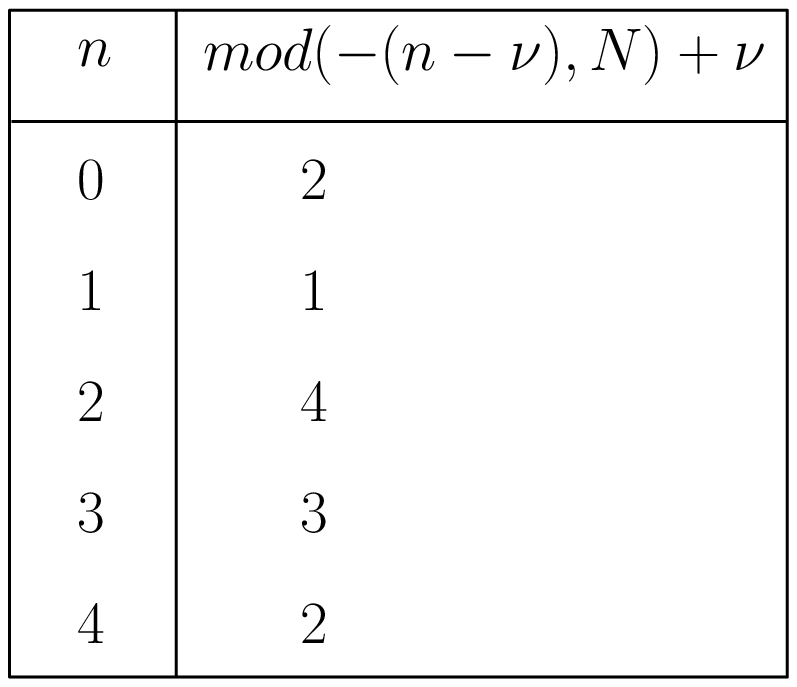}} \quad \quad \quad
  \subfloat[]{\includegraphics[width=.55\textwidth]{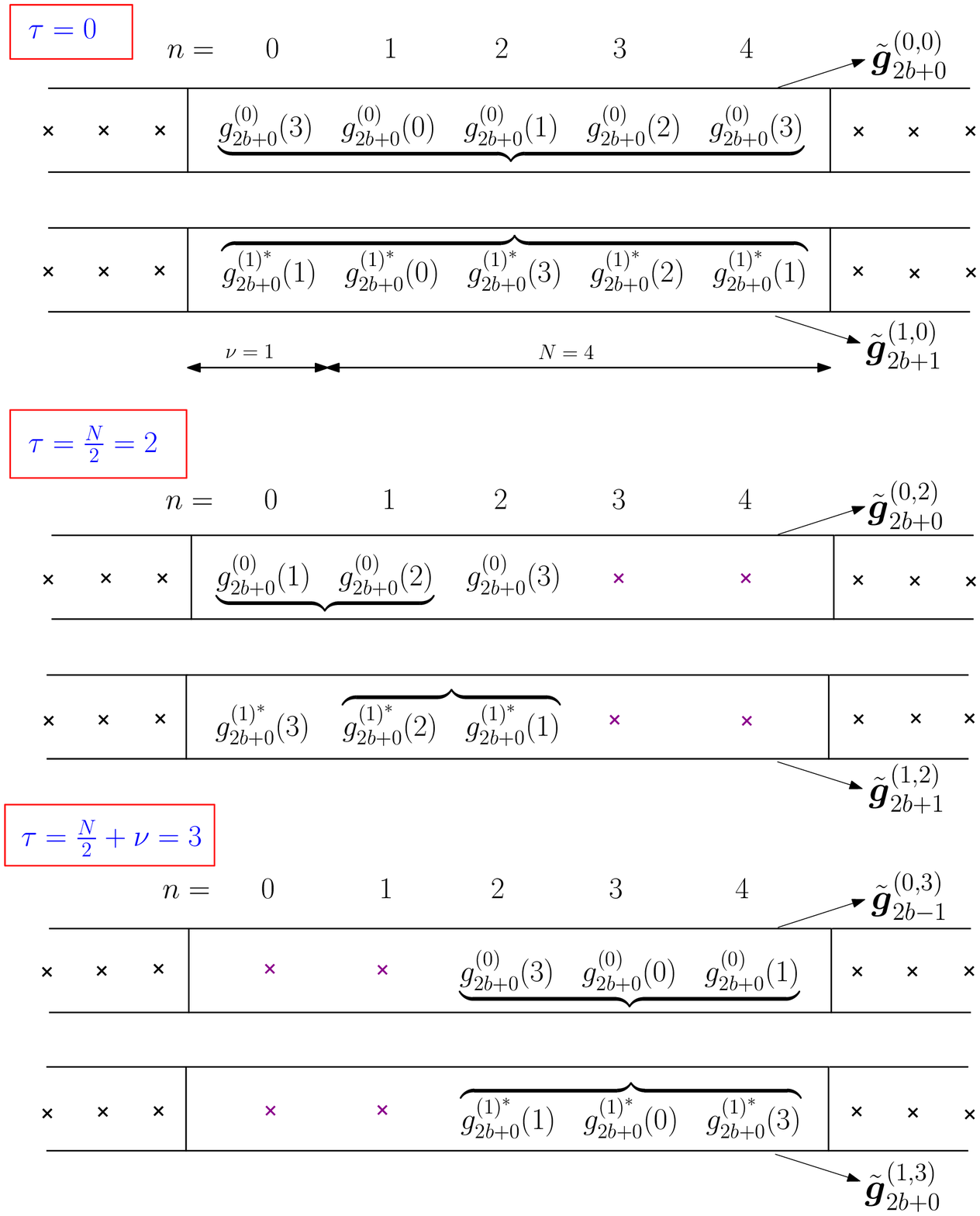}}
  \caption{Illustration of the cross-correlation  between the $(N+\nu)$-length blocks, with $N=4$ and $\nu=1$, and for $\tau=0, \frac{N}{2}, \frac{N}{2}+\nu$.}
  \label{fig:frames}
\end{figure*}

\vspace{0.25cm}
Illustrative examples for \textit{Proposition 1} are provided  in Fig. \ref{fig:frames} for the AL-OFDM signal with $N=4$, $\nu=\frac{N}{4}=1$, $\tau=0$, $\tau=2$ $(=\frac{N}{2})$, and $\tau=3$ $(=\frac{N}{2}+\nu)$. Note that the vector components are written based on (\ref{eq:Gm11})-(\ref{eq:Gm22}), given in the appendix, and by taking into account the relationship between  $\tilde{\boldsymbol{g}}^{(f)}_{2b+u}$ and $\boldsymbol{g}^{(f)}_{2b+u}$.
The uncorrelated and correlated samples are indicated by using '$\times$' and braces, respectively.

Based on results of \textit{Proposition 1}, we define the following cross-correlation

\vspace{-0.25cm}
\begin{equation}
\begin{array}{ll}
R_g(\tau)=& {\rm{E}}\left\{ \boldsymbol{\tilde{g}}_{q}^{(0,\tau)}\left[\boldsymbol{\bar{{g}}}_{q+1}^{(1,\tau)}\right]^{T}\right\} \\
& \hspace{-0.6cm}\triangleq \displaystyle \lim_{N_{B}\rightarrow\infty}\frac{1}{N_{B}}\sum_{q=0}^{N_{B}-1} \boldsymbol{\tilde{g}}_{q}^{(0,\tau)}\left[\boldsymbol{\bar{{g}}}_{q+1}^{(1,\tau)}\right]^{T},
\end{array}
\label{eq:E_y22}
\end{equation}
where $\textrm{E}\{.\}$ indicates the statistical expectation over the block,
$\boldsymbol{\bar{g}}_{q+1}^{(1,\tau)}$ is an $(N+\nu)$-length block with components
 $\bar{g}_{q+1}^{(1,\tau)}(p)= \tilde{g}_{q+1}^{(1,\tau)}(mod(-(p- \nu),N)+ \nu)$, $p=0,1,...,N+\nu-1$,  the superscript $T$ denotes matrix transpose, and $N_B$ is the number of blocks.

By using \textit{Proposition 1}, one can easily see that for $\tau=0,1,...,N+\nu-1$, the cross-correlation for AL-OFDM and SM-OFDM signals is respectively given by

\begin{equation}
R^{\textrm{AL}}_g(\tau) =\left\{ \begin{array}{cc}
\frac{1}{2}(N+\nu)\sigma_{d}^{2}, & \tau=0,\\
\frac{1}{2}(\nu+1)\sigma_{d}^{2}, & \tau=\frac{N}{2},\\
\frac{1}{2}(2\nu+1)\sigma_{d}^{2}, & \tau=\frac{N}{2}+\nu,\\
0, & \textrm{otherwise},
\end{array}\right.\label{eq:E_y}
\end{equation}
and
\begin{equation}
R^{\textrm{SM}}_g(\tau) =0,\label{eq:E_SM}
\end{equation}
where $\sigma_{d}^{2}$ is the variance of the modulated symbols\footnote{Note that based on the Parseval's theorem, the variance of the modulated symbols is equal to the variance of the samples in the block $\boldsymbol{g}^{(f)}_{Ub+u}$  at the output of the IFFT.}. Note that the factor $\frac{1}{2}$ in (\ref{eq:E_y}) is due to the fact that
 correlation exists only between the $(N+\nu)$-length blocks which belong to the same AL block.
According to (\ref{eq:E_y}) and (\ref{eq:E_SM}), 
$R_g(\tau)$ provides a feature for the identification of the AL-OFDM and SM-OFDM signals.

\begin{figure*}
 \hspace{-1cm} \subfloat[]{\includegraphics[width=.55\textwidth]{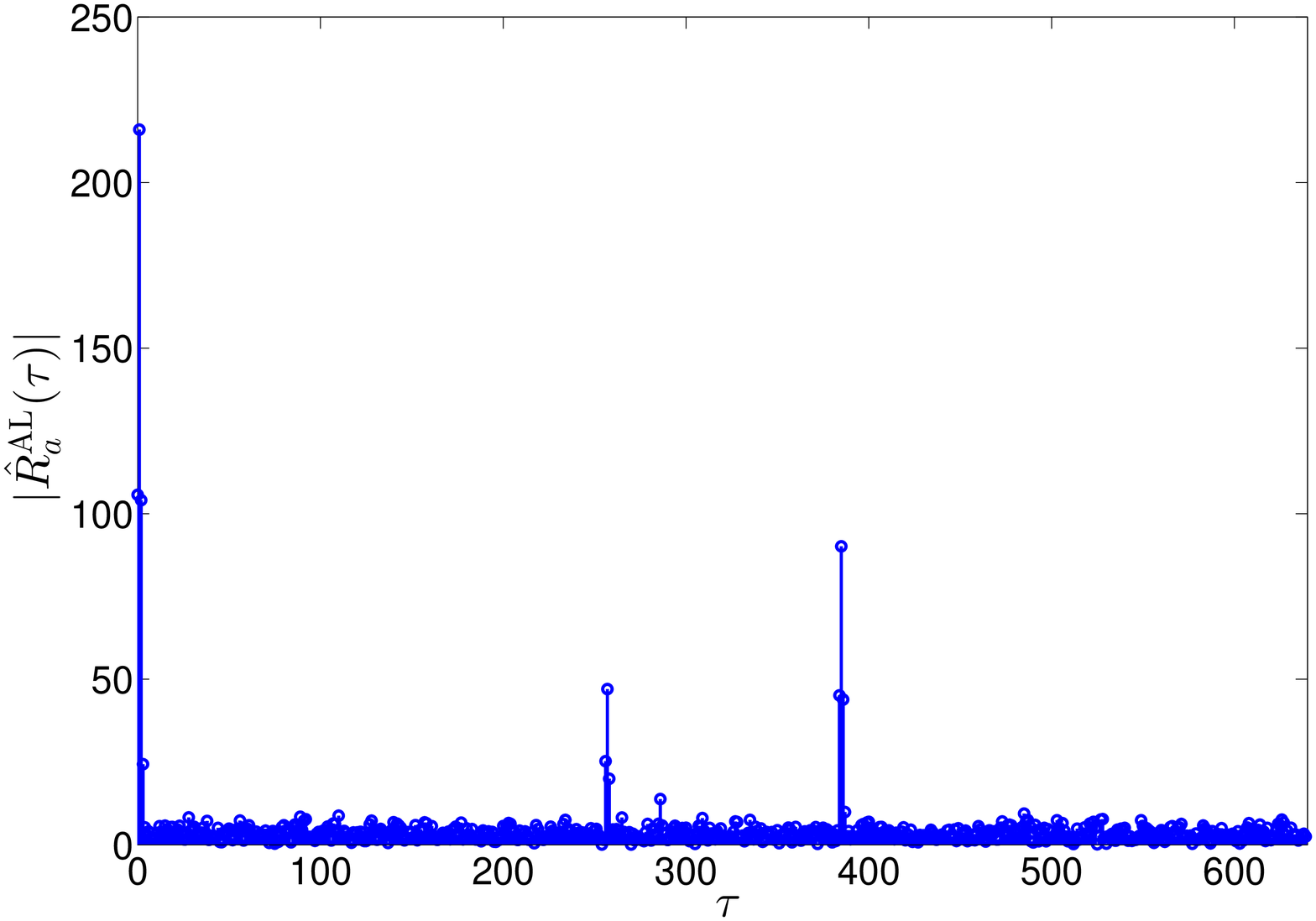}}
  \subfloat[]{\includegraphics[width=.55\textwidth]{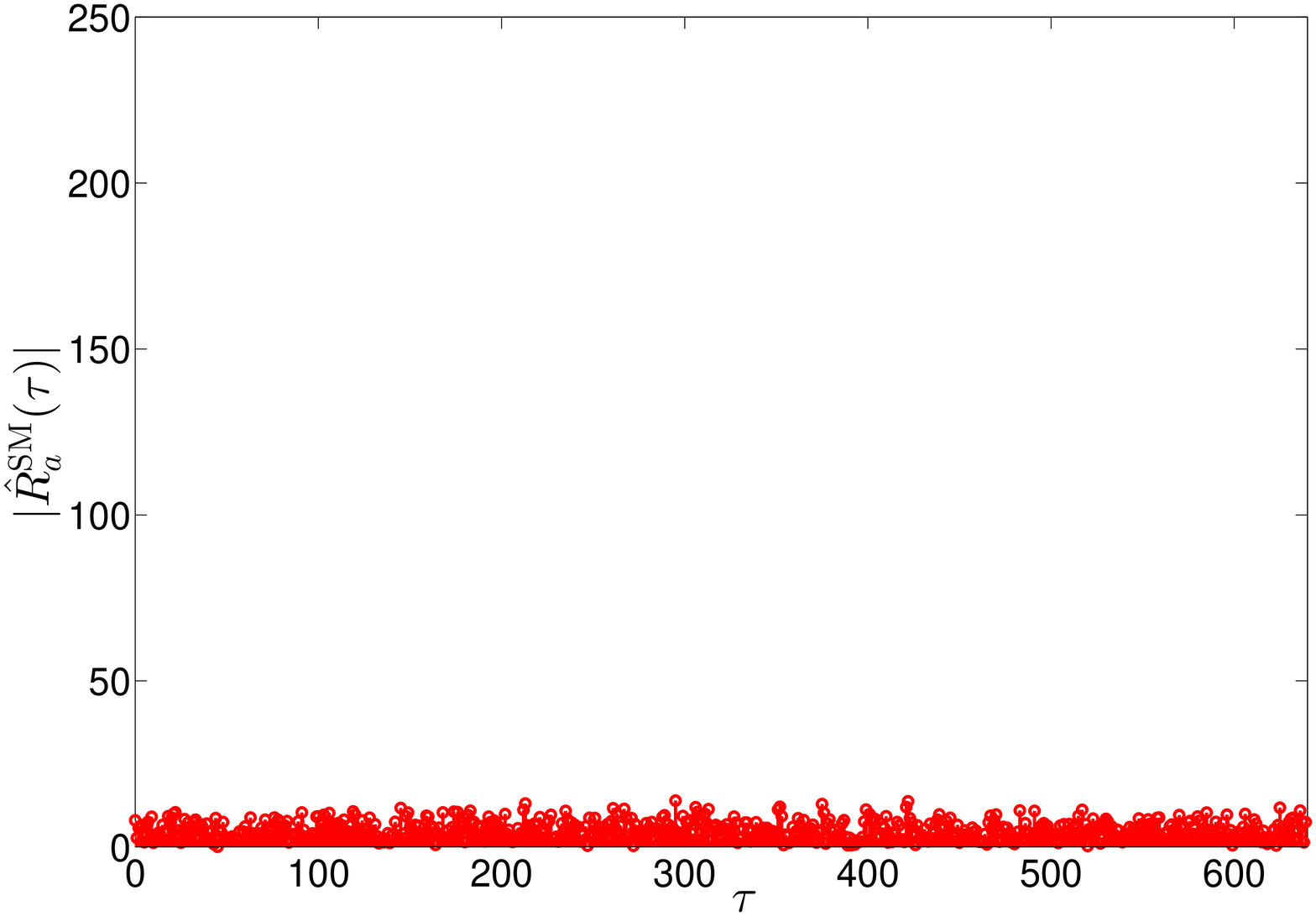}}
  \caption{$|\hat{R}_a(\tau)|$ with QPSK modulation, $N=512$, $\nu=N/4$, and $N_B=100$, at
SNR = 10 dB over multipath Rayleigh fading channel, $L_h=4$, for (a) AL-OFDM and (b) SM-OFDM signals.}
  \label{fig:AL-SM-OFDM}
\end{figure*}

 \vspace{0.4cm}
\underline{\textit{
Receive-side}}

Without loss of generality, we assume that the first intercepted sample corresponds to the start of an OFDM block; later in the paper, we will relax this assumption. Let us define the sequence $\boldsymbol {r}^{(i,\tau)}$, whose components are given by $r^{(i,\tau)}(k)=r^{(i)}(k+\tau)$, $\tau=0,1,...,N+\nu-1$, and further divide it into
$(N+\nu)$-length\footnote{We assume that the OFDM block length is known. Different algorithms in the literature, e.g., \cite{punchihewa2011blind}, can be combined with the proposed algorithm to blindly estimate the OFDM block length.} blocks, i.e.,
$\boldsymbol{r}^{(i,\tau)}=[..., \boldsymbol{a}_{-1}^{(i,\tau)}, \boldsymbol{a}_{0}^{(i,\tau)},\boldsymbol{a}_{1}^{(i,\tau)},...] $,
where $\boldsymbol{a}_{q}^{(i,\tau)}=[a_q^{(i,\tau)}(0), . . . ,a_q^{(i,\tau)}(N+\nu-1)]$, with $a_{q}^{(i,\tau)}(p)=r^{(i,\tau)}(q(N+\nu)+p),\; \;p=0,1,...,N+\nu-1.$

By using  (\ref{eq:RX_signal}), the definition of the correlation in (\ref{eq:E_y22}), (\ref{eq:E_y}),  and (\ref{eq:E_SM}), and taking into account the independence between the transmitted data symbols, noise, and channel coefficients, for $\tau=0,1,...,N+\nu-1$ it is straightforward to find that

\begin{equation}
\begin{array}{ll}

R_a^{\textrm{AL}}(\tau)= & {\rm{E}}\left\{ \boldsymbol{\tilde{a}}_{q}^{(0,\tau)}\left[\boldsymbol{\bar{{a}}}_{q+1}^{(1,\tau)}\right]^{T}\right\} \\  &\hspace{-0.85cm} =\left\{ \begin{array}{ll}
\frac{\sigma_{d}^{2}}{2} (N+\nu) \Xi(\tau), & \tau=0,1,...,L_h-1,\\ & \\
\frac{\sigma_{d}^{2}}{2} (\nu+1) \Xi(\tau), &  \tau=\frac{N}{2},\frac{N}{2}+1,...,\\ & \frac{N}{2}+L_h-1,\\
\frac{\sigma_{d}^{2}}{2} (2\nu+1)  \Xi(\tau), &  \tau=\frac{N}{2}+\nu,\frac{N}{2}+\nu+1,\\ & ...,\frac{N}{2}+\nu+L_h-1,\\ & \\
0, & \textrm{otherwise},\end{array}\right.

\end{array}
\label{eq:R_AL}\end{equation}
and
\begin{equation}
R_a^{\textrm{SM}}(\tau)=0,
\label{eq:R_SM}\end{equation}
where $\Xi(\tau)=\sum_{l,l'=0}^{L_{h}-1}(h_{00}(l)h_{11}(l')-h_{10}(l)h_{01}(l')) \delta(\tau-l-l')$.

Fig. \ref{fig:AL-SM-OFDM} shows the absolute value of the estimated cross-correlation, $|\hat{R_a}(\tau)|$, $\tau=0,1,...,N+\nu-1$, for both AL-OFDM and SM-OFDM signals with QPSK modulation, $N=512$, $\nu=N/4$, and $N_B=100$ over multipath Rayleigh fading channel with $L_h=4$ at SNR=10 dB. Note that the limited observation period results in  non-zero, but statistically non-significant values  for $|R_a^{\textrm{AL}}(\tau)|$ and $|R_a^{\textrm{SM}}(\tau)|$ at the null positions.
The existence of the statistically significant peaks in $|R_a^{\textrm{AL}}(\tau)|$ will be used as a discriminating feature to identify AL-OFDM and SM-OFDM signals.   
It is worthy to mention that the first received sample does not have to correspond to the start of an OFDM block. In such a case, the peaks in Fig.  \ref{fig:AL-SM-OFDM} (a) will be cyclically shifted by the number of samples corresponding to the delay between the first received sample and the start of the first received OFDM block, which does not affect the discriminating feature.

\subsection{ Discriminating feature and decision criterion ($N_r=2$ case)}

The identification of AL-OFDM and SM-OFDM signals relies on detecting 
whether statistically significant peaks are present or not in $|\hat{R_a}(\tau)|$, $\tau=0,1,...,N+\nu-1$. This can be formulated as a binary hypothesis testing problem, where under hypothesis $\mathcal{H}_0$ (no peaks are detected) SM-OFDM is decided to be the received signal, whereas AL-OFDM signal is selected under hypothesis $\mathcal{H}_1$ (peaks are detected). Here we propose a statistical test to detect the peak presence.

Without loss of generality, we assume that the number of observed samples, $K$, corresponds to an integer number of OFDM blocks, $N_B=\frac{K}{N+\nu}$ \footnote{If this not the case, zeros can be added after the observed samples to ensure this relation. Additionally, it is worth noting that the number of received blocks used for signal identification, $N_B$, is finite.}. In this case, $R_a(\tau)$ can be estimated as

\begin{equation}
\hat{R}_a(\tau)=\frac{1}{N_{B}}\sum_{q=0}^{N_{B}-1}\boldsymbol{a}_{q}^{(0,\tau)}\left[\boldsymbol{\bar{a}}_{q+1}^{(1,\tau)}\right]^{T}.
\label{eq:Rm_est}\end{equation}

Following \cite{kay1993fundamentals}, $\hat{R}_a(\tau)$ can be represented as

\begin{equation}
\hat{R}_a(\tau)=R_a(\tau)+\psi(\tau),\label{eq:Rm_est2}\end{equation}
where $\psi(\tau)$ is a zero-mean random variable representing the estimation error, which vanishes asymptotically  ($N_B \rightarrow \infty$).
As shown in (\ref{eq:R_AL}), under the assumption that the first received sample corresponds to the start of an OFDM block, $R_a^{\textrm{AL}}(\tau)$ exhibits $L_h$ peaks around $\tau=0, \frac{N}{2}$, and $\frac{N}{2}+\nu$. In general, if the first received sample corresponds to the $\tau_0$th point in the OFDM block, the peaks in $R_a^{\textrm{AL}}(\tau)$ will be around $\tau=\tau_0$, $\tau=\tau_1=mod(\tau_0+\frac{N}{2}, N+\nu)$, and $\tau=\tau_2=mod(\tau_0+\frac{N}{2}+\nu, N+\nu)$.

Based on (\ref{eq:R_AL}), (\ref{eq:Rm_est2}) can be written for the AL-OFDM  signal  as

\vspace{-0.25cm}
\begin{equation}
\hat{R}_a^{\textrm{AL}}(\tau)= {R}_a^{\textrm{AL}}(\tau)+\psi^{\textrm{AL}}(\tau),
\label{eq:Rm_AL2}
\end{equation}
where ${R}_a^{\textrm{AL}}(\tau)$ is non-zero for $\tau\in \Omega_0$, $\Omega_{0}=\{\tau_{0},\tau_{0}+1,...,\tau_{0}+L_{h}-1\}\cup\{\tau_{1},\tau_{1}+1,...,\tau_{1}+L_{h}-1\} \cup\{\tau_{2},\tau_{2}+1,...,\tau_{2}+L_{h}-1\}$. 

Furthermore, based on (\ref{eq:R_SM}), (\ref{eq:Rm_est2}) can be written for the SM-OFDM signal as

\vspace{-0.25cm}
\begin{equation}
\hat{R}_a^{\textrm{SM}}(\tau)=\psi^{\textrm{SM}}(\tau), \; \forall \tau=0,1,...,N+\nu-1.\label{eq:Rm_SM}\end{equation}

As such, if $R_a(\tau) \neq 0$ \footnote{Henceforth, the superscript AL or SM is dropped in the cross-correlation, as this is not known at the receive-side.} for at least one value of $\tau$, the AL-OFDM signal is declared present ($\mathcal{H}_1$ is true); otherwise, the  SM-OFDM signal is declared present ($\mathcal{H}_0$ is true). The  proposed statistical test detects the presence of the non-zero value of ${R}_a(\tau)$ as follows. For $\tau=0,1,...,N+\nu-1$, we define $\tau_p$ as the value of $\tau$ that maximizes $|\hat{R}_a(\tau)|$,
 
\begin{equation}
\tau_{p}=\arg\max_{\tau}|\hat{R}_a(\tau)|.\label{eq:mp}
\end{equation}

Based on the results provided in (\ref{eq:R_AL}), one can notice that for the AL-OFDM signal, $\tau_p$ will take values in the set $\{\tau_{0},\tau_{0}+1,...,\tau_{0}+L_{h}-1\}$. Depending on the $\tau_p$ value within this range, in order to eliminate all possible peak positions, we consider the set 
$\Omega_{p}=\{ \tau_{p}-L_{h}+1,...,\tau_{p},...,\tau_{p}+L_{h}-1\}$ $\cup\{ \tau_{p1}-L_{h}+1,...,\tau_{p1},...,\tau_{p1}+L_{h}-1\}\cup\{ \tau_{p2}-L_{h}+1,...,\tau_{p2},...,\tau_{p2}+L_{h}-1\} $, with $\tau_{p1}=mod(\tau_p+\frac{N}{2}, N+\nu)$ and $\tau_{p2}=mod(\tau_p+\frac{N}{2}+\nu, N+\nu)$. As such, $R_a(\tau)=0$ for both AL-OFDM and SM-OFDM signals for the delay range $\tau\notin \Omega_p, \; \tau=0,1,...,N+\nu-1$. This result will be used in the definition of the test statistic to avoid the statistically significant peaks. 

When the SM-OFDM signal is received (under hypothesis $\mathcal{H}_0$), $\hat{R}_a(\tau)=\psi(\tau)$ has an asymptotic complex Gaussian distribution with zero-mean and variance $\sigma^2$ \cite{kay1993fundamentals,brillinger2001time}. Therefore, 
the normalized cross-correlation, $\sqrt{\frac{2}{\sigma^{2}}}\hat{R}_a(\tau)$,  asymptotically follows a complex Gaussian distribution with zero-mean and variance equal to 2. Based on that, we define the  function $\mathcal{F}(\tau)$ as    

\vspace{-0.25cm}
\begin{equation}
\mathcal{F}(\tau)=\frac{2|\hat{R}_a(\tau)|^2}{\frac{1}{N+\nu-\rttensor{\Omega}_p}{\textstyle {\displaystyle \sum_{\tau'\notin\Omega_{p}}}|\hat{R}_a(\tau')|^2}},\label{eq:test}\end{equation}
where $\rttensor{\Omega}_p$ is  the cardinality of the set $\Omega_p$.\footnote{Note that in a practical implementation of the algorithm, knowledge of $L_h$ is not required; a reasonably large value is considered. However, this is significantly low when compared to $N+\nu$ and does not affect the algorithm performance.} Note that the denominator in (\ref{eq:test}) is an estimate of the variance of $\hat{R}_a(\tau)$ under hypothesis $\mathcal{H}_0$, which converges to $\sigma^2$ when $N$ goes to infinity.  As such, $\mathcal{F}(\tau)$   has an asymptotic chi-square distribution with two degrees of freedom under hypothesis $\mathcal{H}_0$ \cite{papoulis2001probability}. Accordingly, we define the test statistic $\Upsilon$ as  

\vspace{-0.2cm}
\begin{equation}
\Upsilon=\max \mathcal{F}(\tau), \quad \tau=0,1,...,N+\nu-1.\label{eq:test1}
\end{equation}

Then we set a threshold, $\eta$, to yield a desired probability of false alarm, $P_{fa}$, i.e., $P_{fa}=P(\mathcal{H}_1 | \mathcal{H}_0)=P (\Upsilon \geq \eta)$. 
Using the expression of  the cumulative distribution function (CDF) of the chi-square distribution with two degrees of freedom \cite{papoulis2001probability}, we can find that

\begin{equation}
P(\Upsilon<\eta)=(1-e^{\frac{-\eta}{2}})^{(N+\nu)}.\label{eq:P_max2}\end{equation}

Since $P_{fa}=1-P(\Upsilon<\eta)$, the threshold, $\eta$, can be calculated for a given $P_{fa}$ as

\begin{equation}
\eta=-2\ln(1-(1-P_{fa})^{\frac{1}{N+\nu}}).\label{eq:threshold}\end{equation} 

Finally, if $\Upsilon\geq\eta$, the AL-OFDM signal is decided to be  received; otherwise, the SM-OFDM signal is selected. A summary of the proposed identification algorithm is given~below.

\floatname{algorithm}{}
\begin{algorithm}
\renewcommand{\thealgorithm}{}
\caption{\textbf{\hspace{-0.25cm}Summary of the proposed identification algorithm ($N_r=2$)}}
\begin{algorithmic}[0]
\State \small{\textbf{Required signal pre-processing:} Estimation of the   OFDM block length ($N+\nu$).}
\State \textbf{Input:} The observed $K$ samples from two receive antennas $\left\{ r^{(0)}(k)\right\} _{k=0}^{K-1}$ and $\left\{ r^{(1)}(k)\right\} _{k=0}^{K-1}$.

\State - Estimate the cross-correlation $R_a(\tau)$, $\tau=0,1,...,N+\nu-1$,  using (\ref{eq:Rm_est}).
\State - Compute $\Upsilon$ using (\ref{eq:test}) and (\ref{eq:test1}).
\State - Compute $\eta$ using (\ref{eq:threshold}) based on the target $P_{fa}$.
\If {$\Upsilon \geq \eta$}
\State - the AL-OFDM signal is declared present ($\mathcal{H}_{1}$ true).
\Else
\State - the SM-OFDM signal is declared present ($\mathcal{H}_{0}$ true).
\EndIf

\end{algorithmic}
\end{algorithm}

\subsection{Discriminating feature and decision criterion ($N_r > 2$ case)}

In the previous section we considered  two receive antennas ($N_r=2$); here, we  generalize the proposed identification algorithm to $N_r >2$. Basically, the cross-correlations between each pair of the receive antennas will be combined to improve the discriminating feature. Similar to (\ref{eq:Rm_est}), the cross-correlation between the $i$th and $j$th receive antennas, $\hat{R}_{a,i,j}(\tau), i=0,1,...,N_r-2,$ $j= i+1, i+2, ..., N_r-1$, can be estimated as

\vspace{-0.25cm}
\begin{equation}
\hat{R}_{a,i,j}(\tau)=\frac{1}{N_{B}}\sum_{q=0}^{N_{B}-1}\boldsymbol{a}_{q}^{(i,\tau)}\left[\boldsymbol{\bar{a}}_{q+1}^{(j,\tau)}\right]^{T}.
\label{eq:Rm_est22}\end{equation}

For each pair of receive antennas, the  function $\mathcal{F}_{i,j}(\tau)$, $\tau=0,1,...,N+\nu-1$, is calculated~as

\vspace{-0.1cm}
\begin{equation}
\mathcal{F}_{i,j}(\tau)=\frac{2|\hat{R}_{a,i,j}(\tau)|^2}{\frac{1}{N+\nu-\rttensor{\Omega}_{p,i,j}}{\textstyle {\displaystyle \sum_{\tau\notin\Omega_{p,i,j}}}|\hat{R}_{a,i,j}(\tau)|^2}},\label{eq:test22}\end{equation}
and the  functions for all pairs of receive antennas are combined as

\vspace{-0.35cm}
\begin{equation}
\mathcal{F}_{c}(\tau)=\sum_{i=0}^{N_{r}-2}\sum_{j=i+1}^{N_{r}-1}\mathcal{F}_{i,j}(\tau).\label{eq:Tc}
\end{equation}

Accordingly, the test statistic is defined as

\vspace{-0.25cm}
\begin{equation}
\Upsilon=\max \mathcal{F}_c(\tau).\label{eq:test12}
\end{equation}

As $\mathcal{F}_{i,j}(\tau)$ has an asymptotic chi-square distribution with two degrees of freedom under hypothesis $\mathcal{H}_0$, $\mathcal{F}_{c}(\tau)$ asymptotically follows the chi-square distribution with $2N_c$ degrees of freedom,
where $N_c=\frac{N_r(N_r-1)}{2}$ is the number of the pairs of receive antennas.
Hence, for a certain $P_{fa}=P(\mathcal{H}_1 | \mathcal{H}_0)=P(\Upsilon \geq \eta)$ we set the threshold  based on the CDF of this chi-square distribution, i.e.,

\vspace{-0.25cm}
\begin{equation}
(1-P_{fa})^{\frac{1}{N+\nu}}=\frac{\gamma(N_{c},\eta/2)}{(N_c-1)!},\label{eq:new_th}
\end{equation}
where $\gamma(\centerdot, \centerdot)$ is the lower incomplete Gamma function \cite{simon2007probability}. Note that for $N_r=2$, the threshold, $\eta$, in (\ref{eq:new_th}) can be expressed as in (\ref{eq:threshold}). On the other hand, for $N_r >2$, the threshold $\eta$ cannot be expressed in a closed form; in such cases, this is numerically calculated  for a certain $P_{fa}$  using the bisection method \cite{stoer2002introduction}.

\subsection{Computational complexity}

The computational complexity of the proposed algorithm is measured by the required number of floating point operations (flops) \cite{watkins2002fundamentals}, which can be easily found to be equal to
$N_c(6N_B(N+\nu)^2+(2N_B+4)(N+\nu))$. For example, with $N=256$, $\nu=\frac{N}{4}$, $N_r=2$, and $N_B=100$, the proposed algorithm requires 61,505,280 flops. Practically speaking, a microprocessor with 79.992 Giga-flops\footnote{[online], available: http://download.intel.com/support/processors/corei7/sb/core$\_$i7-900$\_$d.pdf} can perform the calculations needed for the proposed algorithm in approximately 769 $\mu$sec.

\section{Simulation results \label{sec:simulation}}

\subsection{Simulation setup}

The identification performance of the proposed algorithm was evaluated using Monte Carlo simulations with 1000 trials for each signal type. The OFDM signals are generated based on the IEEE 802.11e standard, with a useful OFDM block duration of 91.4 $\mu$sec and a subcarrier spacing of $10.94$ kHz.
Unless otherwise mentioned, the  modulation was QPSK, the number of OFDM subcarriers $N=256$ (2.5 MHz double-sided bandwidth), the cyclic prefix $\nu=N/4$, the number of observed OFDM blocks $N_{B}=100$, the number of receive antennas $N_r=2$,  and the probability of false alarm $P_{fa}=10^{-3}$. 
Furthermore, the received signal was affected by a frequency selective Rayleigh fading channel\footnote{While a Rayleigh fading channel is considered here, it is worth noting that a similar performance is achieved under multipath Nakagami-$m$ fading conditions, as the distribution of the test statistic is similar under diverse channel conditions, as shown by~simulations.} consisting of $L_h=4$ statistically independent taps, with an exponential power delay profile \cite{patzold2002methods}, $\sigma^{2}(l)=\exp(-l/5)$, where $l=0,...,L_h-1$. A Butterworth filter was used at the receive-side to remove the out-of-band noise, and the SNR was considered at the output of this filter. The average probability of correct identification, $P_c=0.5(P(\lambda=\textrm{AL}|\textrm{AL})+P(\lambda=\textrm{SM}|\textrm{SM}))$, was employed as a performance measure, where $\lambda$ is the estimated signal~type. 
  
\vspace{-0.2cm}
\subsection{Performance evaluation}

\begin{figure}
\begin{centering}
\includegraphics[width=0.75\textwidth]{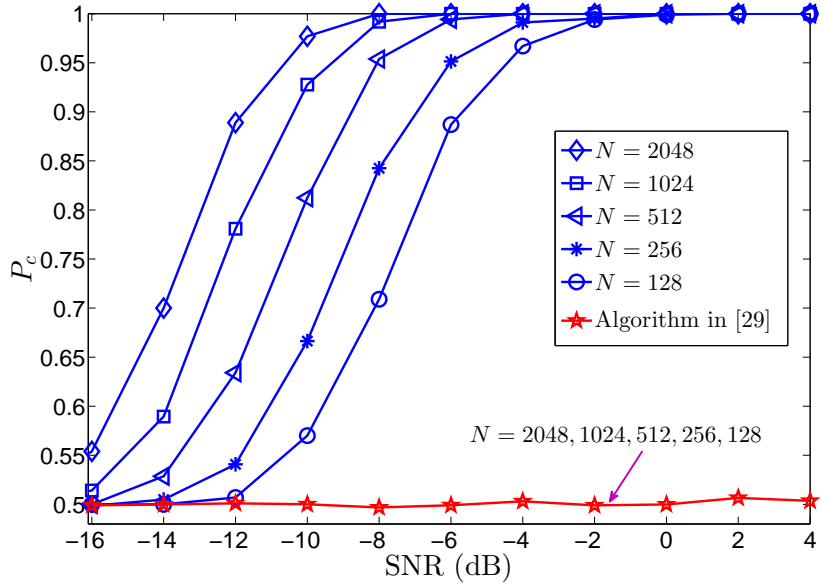}
\par\end{centering}
\caption{Performance comparison between the proposed algorithm  and the one in \cite{MareyICC2013} for various numbers of OFDM subcarriers, $N$, with $N_B=100$.\label{fig:N_effect}}
\end{figure}

Fig. \ref{fig:N_effect} shows the performance of the proposed algorithm in comparison with that in \cite{MareyICC2013} for different numbers of OFDM subcarriers, $N$. Apparently, the proposed algorithm outperforms the algorithm in \cite{MareyICC2013}, which basically fails; the reason is that the latter requires a large number of OFDM blocks  to estimate the discriminating feature, e.g., simulation results show that $N_B=10,000$ is needed to reach $P_c \approx 1$ at SNR = -4 dB. 

In terms of computational complexity, the algorithm in \cite{MareyICC2013} requires $(N+\nu)(8N_B+4)$ flops. If we compare the complexity of this algorithm and the proposed algorithm for given values of $N_B$, $N$, $\nu$, and $N_r$, the algorithm in \cite{MareyICC2013} is less computationally demanding. For example, for $N_B=100$, $N=256$, $\nu=N/4$, and $N_r=2$, the former requires 257,280 flops, while the latter needs 61,505,280 flops. However, such a complexity comparison is not fair due to the difference in performance (as discussed above, based on results in Fig. 5). If we consider the $N_B$ values for which the algorithms reach $P_c \approx 1$ at a given SNR, along with the fact that the time to make a decision consists of both observation and computing times, then it can be easily found that the algorithm in \cite{MareyICC2013} requires a longer time for decision. For example, when a microprocessor with 79.992 Giga-flops is employed for computation, the algorithm in [29] needs 1.1428 sec to make a decision with $P_c \approx 1$ at SNR= -4 dB ($N_B=10,000$), whereas the proposed algorithm requires only 12.194 $m$sec ($N_B=100$).

Furthermore, it can be observed from  Fig. \ref{fig:N_effect} that the identification performance of the proposed algorithm significantly improves  by increasing $N$.
This is because the peak values in $|R_a^{\rm{AL}}(\tau)|$ are significantly enhanced, i.e., $|R_a^{\rm{AL}}(\tau)|$ is proportional to $(N+\nu)$ as can be noticed from (\ref{eq:R_AL}). This  reflects on the discriminating feature and leads to identification performance improvement.

\subsection{Effect of the number of OFDM blocks}

\begin{figure}
\begin{centering}
\includegraphics[width=0.75\textwidth]{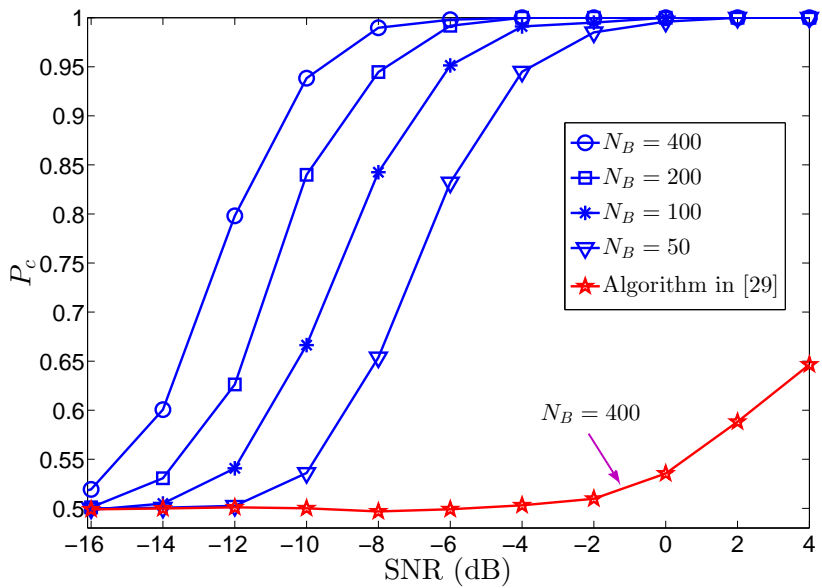}
\par\end{centering}
\caption{The effect of the number of OFDM blocks, $N_B$, on the average probability of correct identification, $P_{c}$. \label{fig:NB_effect}}
\end{figure}

Fig. \ref{fig:NB_effect} shows the effect of the number of OFDM blocks, $N_B$, on the average probability of correct identification, $P_c$. A comparison with the  algorithm  in \cite{MareyICC2013} for $N_B=400$ is also included. As expected, increasing $N_B$ enhances the performance of the proposed algorithm, as it leads to a better estimate of the cross-correlation, $\hat{R}_a(\tau)$. Note that the proposed algorithm provides an excellent performance ($P_c \approx 1$) at SNR = 0 dB and with a small number of blocks, $N_B=50$, whereas the algorithm in \cite{MareyICC2013} does not achieve a good performance even for $N_B=400$. 

\vspace{-0.2cm}
\subsection{Effect of the cyclic prefix length}

\begin{figure}
\begin{centering}
\includegraphics[width=0.75\textwidth]{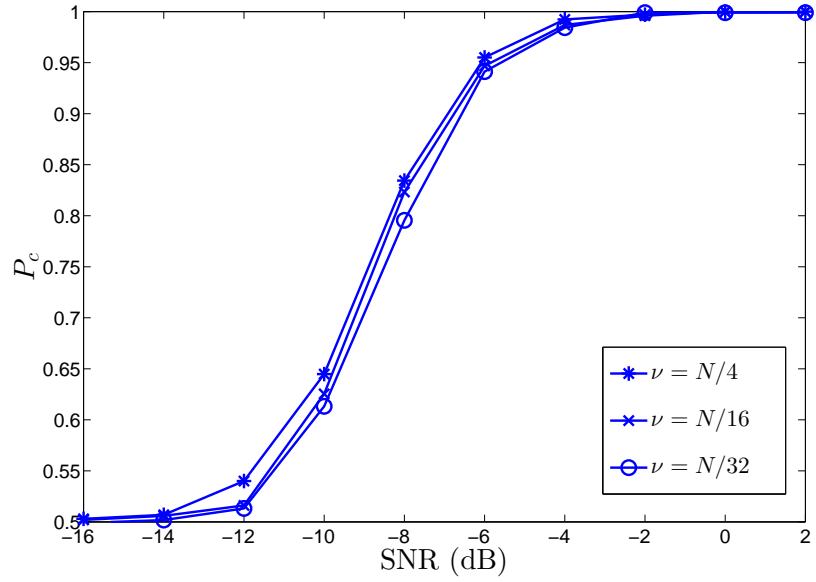}
\par\end{centering}
\caption{The effect of the cyclic prefix length, $\nu$, on the average probability of correct identification, $P_{c}$.  \label{fig:CP}}
\end{figure}

Fig. \ref{fig:CP} shows  the average probability of correct identification, $P_{c}$, for $\nu=N/4$, $N/16$, and $N/32$.
One can notice that the performance  slightly improves  by increasing $\nu$; this is because under the
$\mathcal{H}_1$ hypothesis (the AL-OFDM signal), the peak values in $|\hat{R}_a^{\textrm{AL}}(\tau)|$  slightly increase with $\nu$. 
It is worth noting that the improvement obtained by increasing $N$ is more significant, as was seen in Fig. \ref{fig:N_effect}.

\vspace{-0.2cm}
\subsection{Effect of the number of receive antennas}

\begin{figure}
\begin{centering}
\includegraphics[width=0.75\textwidth]{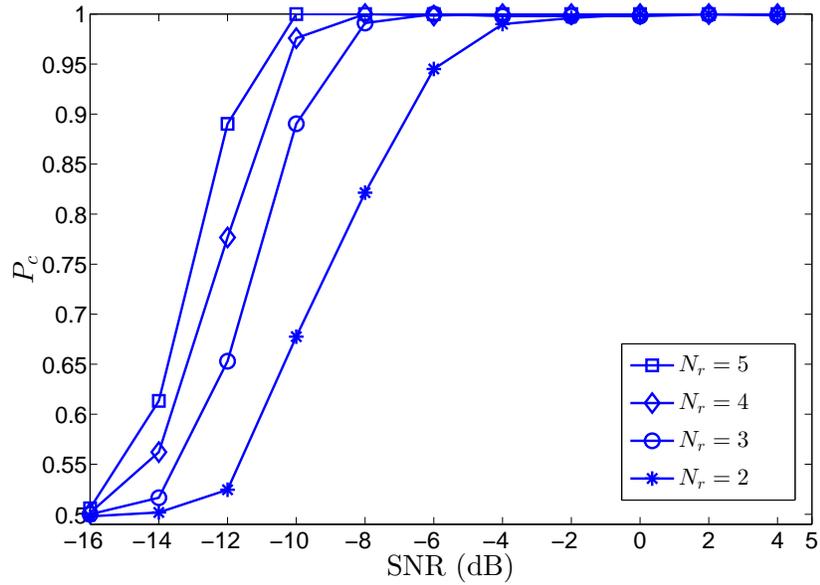}
\par\end{centering}
\caption{The effect of the number of receive antennas, $N_r$, on the average probability of correct identification, $P_{c}$. \label{fig:Nr_effect}}
\end{figure}

Fig. \ref{fig:Nr_effect} illustrates the effect of the number of receive antennas, $N_r$, on the average probability of correct identification, $P_c$. It can be seen that the identification performance is improved by increasing $N_r$.  
For example, with $N_r=5$, an excellent performance is obtained at SNR~$=-10$ dB, when compared with SNR $=-2$ dB for $N_r=2$. However, the computational complexity increases by a factor of 10, according to results presented in Section III.D.

\vspace{-0.2cm}
\subsection{Effect of the modulation format}
\begin{figure}
\begin{centering}
\includegraphics[width=0.75\textwidth]{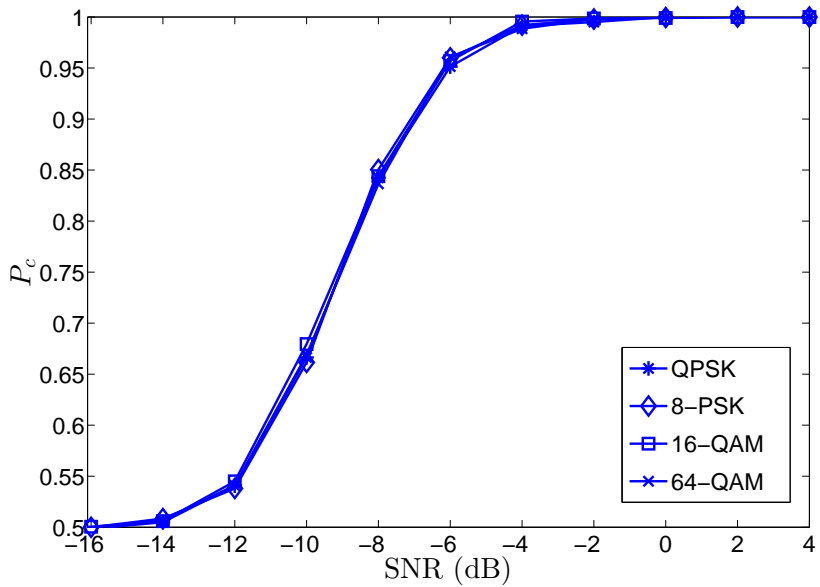}
\par\end{centering}
\caption{The effect of the modulation format on the average probability of correct identification, $P_c$. 
\label{fig:mod_effect}}
\end{figure}

Fig. \ref{fig:mod_effect} presents the effect of the modulation format on the average probability of correct identification, $P_c$. Clearly, it does not affect the performance of the proposed algorithm, as the peak values in $|R^{\textrm{AL}}_a(\tau)|$ do not depend on the modulation format, according to (\ref{eq:R_AL}).  

\vspace{-0.2cm}
\subsection{Effect of the timing offset}

\begin{figure}
\begin{centering}
\includegraphics[width=0.75\textwidth]{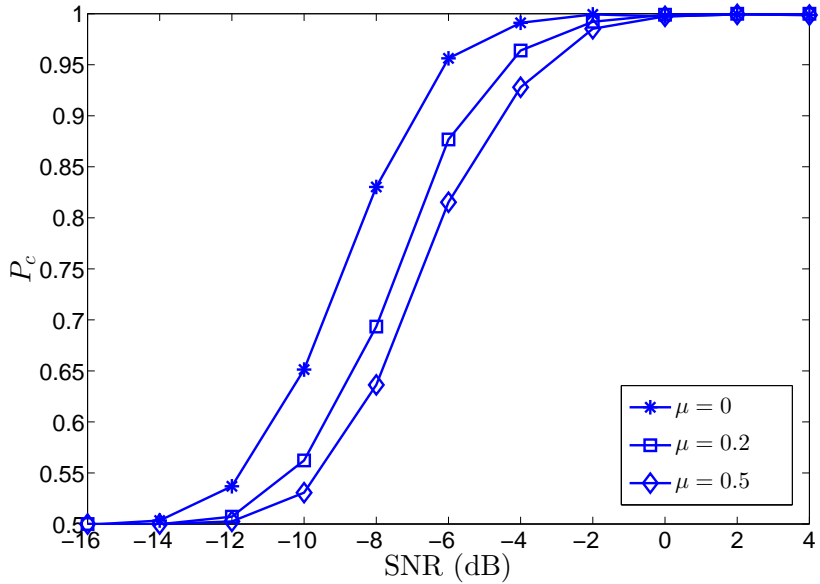}
\par\end{centering}
\caption{The effect of the timing offset on the average probability of correct identification, $P_{c}$. \label{fig:time_offset}}
\end{figure}

Perfect timing synchronization was assumed in the previous study. Here we evaluate the performance of the proposed algorithm in the presence of a timing offset. As mentioned in Section \ref{sec:algorithm}, a timing offset equal to a multiple integer of the sampling period leads to a shift in the positions of the $|R_a^{\textrm{AL}}(\tau)|$ peaks by an amount corresponding to that offset; consequently, this does not affect the discriminating feature. On the other hand, when the timing offset is a fraction of the sampling period, its effect is modeled as a two path channel $[1-\mu, \mu]$, where $0 \leq \mu <1$ is  the normalized timing offset \cite{choqueuse2008hierarchical}. 
Fig. \ref{fig:time_offset} shows the average probability of correct identification, $P_c$, for $\mu=0, 0.2$, and 0.5. The results indicate that while the performance slightly decreases at lower SNRs, it is not affected at higher SNRs. This can be explained, as the effect of $\mu$ can be considered as an additional noise component that affects the peaks in $|R^{\textrm{AL}}_a(m)|$.

\vspace{-0.3cm}
\subsection{Effect of the frequency offset}

\begin{figure}
\begin{centering}
\includegraphics[width=0.75\textwidth]{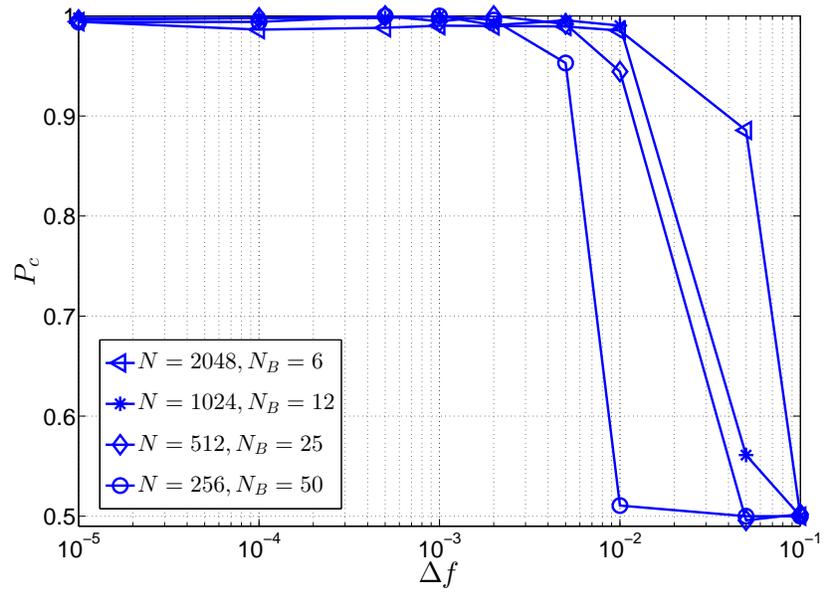}
\par\end{centering}
\caption{The effect of the frequency offset on the average probability of correct identification, $P_{c}$, for different values of $N$ and $N_B$
at SNR = 0 dB.  \label{fig:freq_offset}}
\end{figure}

Fig. \ref{fig:freq_offset} presents the effect of the frequency offset normalized to the subcarrier spacing, $\Delta f$, on the average probability of correct identification, $P_c$, at SNR = 0 dB and for different values of $N$ and $N_B$. 
Note that as the OFDM block duration is constant regardless of $N$ (see Section IV.A), the observation period increases with $N_B$, which leads to an increased effect of the frequency offset on the performance. It is worth noting that a reduced number of OFDM blocks is required to achieve a good performance for a larger number of subcarriers, which results in a lower sensitivity to the frequency offset. Results  in Fig. \ref{fig:freq_offset} show a good robustness for $\Delta f<10^{-2}$ when $N=2048$ and $N_B=6$.

\vspace{-0.2cm}
\subsection{Effect of the Doppler frequency}

\begin{figure}
\begin{centering}
\includegraphics[width=0.75\textwidth]{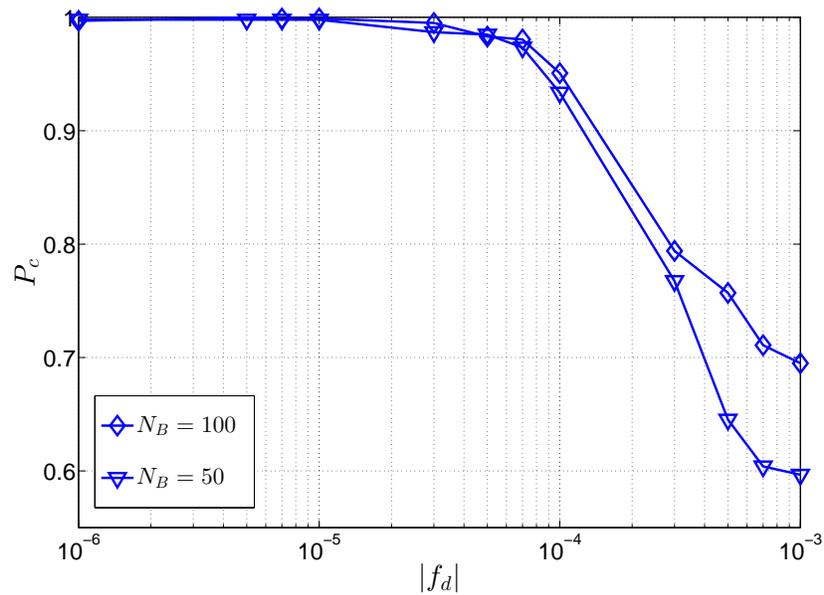}
\par\end{centering}
\caption{The effect of the Doppler frequency on the average probability of correct identification, $P_{c}$, for $N_B=50, 100$
at SNR = 0~dB.  \label{fig:Doppler}}
\end{figure}

The previous analysis assumed constant channel coefficients over the observation period. Here, we consider the effect of the Doppler frequency on the performance of the proposed algorithm.
Fig. \ref{fig:Doppler} shows the average probability of correct identification, $P_c$, versus the absolute value of the Doppler frequency normalized to the sampling rate, $|f_d|$, at SNR $=0$ dB and $N_B=50$ and 100. The results show a good robustness for $|f_d| < 10^{-4}$.   

\vspace{-0.2cm}
\subsection{Effect of the spatially correlated fading}

\begin{figure}
\begin{centering}
\includegraphics[width=0.75\textwidth]{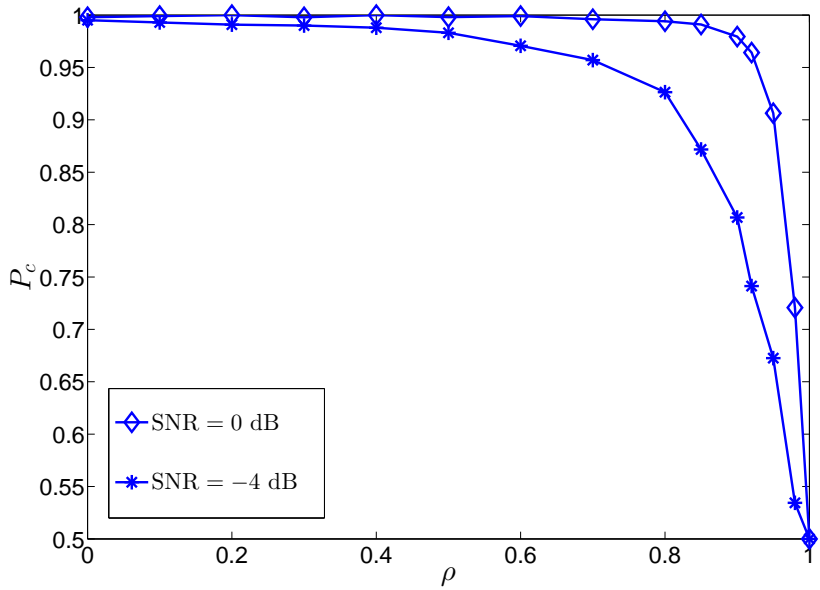}
\par\end{centering}
\caption{The effect of the spatially correlated fading  on the average probability of correct identification, $P_{c}$,
at SNR=-4 dB and 0 dB. \label{fig:spatial_effect}}
\end{figure}

In the previous study, independent fading was considered. Here, we show the effect of the spatially correlated fading on the performance of the proposed algorithm. Fig. \ref{fig:spatial_effect}
shows the average probability of correct identification of the proposed algorithm, $P_c$, versus the spatial correlation coefficient, $\rho$, at SNR $= -4$ dB and 0 dB. As shown in (\ref{eq:R_AL}), the channel coefficients affect the peak values in $|\hat{R}_a^{\textrm{AL}}(\tau)|$ by the factor $|\sum_{l,l'=0}^{L_{h}-1}(h_{00}(l)h_{11}(l')-h_{10}(l)h_{01}(l'))|$. At high values of $\rho$, $h_{00}(l)\thickapprox h_{01}(l)$ and $h_{11}(l')\thickapprox h_{10}(l')$, $l,l'=0,1,...,L_h-1$. As such, the discriminating peaks vanish and the identification performance degrades\footnote{It is worth noting that the same performance is obtained if the spatially correlated fading occurs at the transmit-side; in this case $h_{00}(l)\thickapprox h_{10}(l)$ and $h_{11}(l')\thickapprox h_{01}(l')$, $l,l'=0,1,...,L_h-1$, at high values of the correlation coefficient.}. As expected, the performance is more affected by spatially correlated fading at lower SNR. 
  
\section{Conclusion \label{sec:conclusion}}

The identification of the AL-OFDM and SM-OFDM signals has been investigated in this paper. A new cross-correlation was developed, which provides an efficient  feature for signal identification. Based on the statistical properties of the feature estimate, a novel criterion of decision was introduced. The proposed identification algorithm, which employs the aforementioned  discriminating feature and decision criterion, provides an improved  
 performance when compared with the previous work in the literature, at lower SNR and with reduced observation period. The algorithm has the advantages that it does not require channel and noise power estimation, modulation identification or timing synchronization. Furthermore, it exhibits  a relatively low sensitivity to spatially correlated fading and frequency offset.

\section*{Appendix: Proof of \textit{Proposition 1} }

For the AL-OFDM signal, by using the definition of the ($N+\nu$)-length blocks in the 
$\boldsymbol{s}^{(f,\tau)}$ sequence (see Fig. \ref{fig:g_delay} for the graphical illustration), one can easily express the samples of  $\boldsymbol{\tilde{g}}^{(0,\tau)}_{2b+0}$ and $\boldsymbol{\tilde{g}}^{(1,\tau)}_{2b+1}$ respectively as

\begin{equation}
\tilde{g}_{2b+0}^{(0,\tau)}(n)=\left\{ \begin{array}{ll}
\tilde{g}^{(0)}_{2b+0}(n+\tau), & n=0,1,..., \\ &N+\nu-\tau-1,\\ & \\
\tilde{g}^{(0)}_{2b+1}(n+\tau-N-\nu), & n=N+\nu-\tau, \\ & ...,N+\nu-1,
\end{array}\right.\label{eq:Gm11}
\end{equation}
and 
\begin{equation}
\tilde{g}_{2b+1}^{(1,\tau)}(n')=\left\{ \begin{array}{ll}
\tilde{g}^{(1)}_{2b+1}(n'+\tau), & n'=0,1,..., \\ & N+\nu-\tau-1,\\ & \\
\tilde{g}^{(1)}_{2(b+1)}(n'+\tau-N-\nu), & n'=N+\nu-\tau,\\ & ...,N+\nu-1.
\end{array}\right.\label{eq:Gm21}
\end{equation}

Based on (\ref{eq:IFFT_eq}), for the case of $\tau=0$, it can be written that

\begin{equation}
\begin{array}{l}
\tilde{g}_{2b+0}^{(0,0)}(n)= \tilde{g}_{2b+0}^{(0)}(n) =\frac{1}{\sqrt{N}}\sum_{p=0}^{N-1} c^{(0)}_{2b+0}(p)e^{\frac{j2\pi p(n-\nu)}{N}},  \quad n=0,1,..,N+\nu-1,
\end{array}
\label{eq:app1}\end{equation}
and
\begin{equation}
\begin{array}{l}

\tilde{g}_{2b+1}^{(1,0)}(n')=\tilde{g}_{2b+1}^{(1)}(n')=\frac{1}{\sqrt{N}}\sum_{p=0}^{N-1}c^{(1)}_{2b+1}(p)e^{\frac{j2\pi p(n'-\nu)}{N}},\quad n'=0,1,..,N+\nu-1.
\end{array}
\label{eq:app2}\end{equation}

By using that ${c}_{2b+1}^{(1)}(p)=({c}_{2b+0}^{(0)}(p))^*$, $p=0,1,...,N-1$, for AL-OFDM signal, and taking the complex conjugate of (\ref{eq:app2}), it is straightforward that

\begin{equation}
\begin{array}{l}
\tilde{g}_{2b+1}^{(1,0)^{*}}(n')=\frac{1}{\sqrt{N}}\sum_{p=0}^{N-1}c^{(0)}_{2b+0}(p)e^{\frac{-j2\pi p(n'-\nu)}{N}}, \quad n'=0,1,..,N+\nu-1.
\end{array}
\label{eq:app3}\end{equation}

It is easy to see that  $\tilde{g}_{2b+0}^{(0,0)}(n)=\tilde{g}_{2b+1}^{(1,0)^{*}}(n')$, $n,n'=0,1,...,N+\nu-1$ only when $n'-\nu=mod(-(n-\nu),N)$. A few examples are given as follows: $n=0, n'=2\nu$; $n=\nu, n'=\nu$; $n=\nu+1, n'=N+\nu-1$; and $n=N+\nu-1, n'=\nu+1$.
Hence, one can notice that $n+n'=2\nu$ for $n=0,1,...,\nu$, and $n+n'=N+2\nu$ for $n=\nu+1,..., N+\nu-1$. This leads to the result shown in~(\ref{eq:G_0}).

\vspace{0.15cm}

For $\tau>0$, it is straightforward that $\boldsymbol{\tilde{g}}_{2b+0}^{(0,\tau)}$ and $\boldsymbol{\tilde{g}}_{2b+1}^{(1,\tau)}$ belong to the (same) $b$th AL block for $ n,n'=0,1,...,N+\nu-\tau-1$. 
Moreover, based on the aforementioned results regarding $n$ and $n'$, one can see that $\tilde{g}_{2b+0}^{(0,\tau)}(n)=\tilde{g}_{2b+1}^{(1,\tau)^*}(n'=mod(-(n-\nu),N)+\nu)$ if $n$ and $\tau$ satisfy $n+n' = 2\nu, N+2\nu$ and $n+n'+2\tau = 2\nu, N+2\nu$. 
If $n+n'=2\nu$ and $n+n'+2\tau=N+2\nu$, then $\tau=N/2$, $n=0,1,...,\nu$. This directly leads to the result in (\ref{eq:G_1}). On the other hand, if $n+n'=n+n'+2\tau$ (either equal to $2\nu$ or $N+2\nu$), then $\tau=0$, $n=0,1,...,N+\nu-1$; this leads to the case of $\tau=0$ discussed above. Furthermore,
if $n+n'=N+2\nu$ and $n+n'+2\tau=2\nu$, then $\tau=-N/2$, which is out of range ($0 \leq \tau < N+\nu$).

\vspace{0.15cm}
Moreover, also for the AL-OFDM signal, one can similarly express the samples of 
$\boldsymbol{\tilde{g}}^{(0,\tau)}_{2b-1}$ and $\boldsymbol{\tilde{g}}^{(1,\tau)}_{2b+0}$ respectively as

\begin{equation}
\tilde{g}_{2b-1}^{(0,\tau)}(n)=\left\{ \begin{array}{ll}
\tilde{g}^{(0)}_{2b-1}(n+\tau), & n=0,1,..., \\ & N+\nu-\tau-1,\\ & \\ 
\tilde{g}^{(0)}_{2b+0}(n+\tau-N-\nu), & n=N+\nu-\tau, \\ & ...,N+\nu-1,
\end{array}\right.\label{eq:Gm12}
\end{equation}
and 
\begin{equation}
\tilde{g}_{2b}^{(1,\tau)}(n')=\left\{ \begin{array}{ll}
\tilde{g}^{(1)}_{2b+0}(n'+\tau), & n'=0,1,..., \\ & N+\nu-\tau-1,\\ & \\
\tilde{g}^{(1)}_{2b+1}(n'+\tau-N-\nu), & n'=N+\nu-\tau,\\ & ...,N+\nu-1.
\end{array}\right.\label{eq:Gm22}
\end{equation}

Accordingly, $\tilde{g}_{2b-1}^{(0,\tau)}(n)$ and $\tilde{g}_{2b+0}^{(1,\tau)}(n')$ belong to the (same) $b$th AL block for $n,n'=N+\nu-\tau,...,N+\nu-1$. By using the same analysis as above, one can prove results in (\ref{eq:G_2}).

%

\bibliographystyle{IEEEtran}
\bibliography{IEEEabrv,My_Ref_bib}

\end{document}